# Detritiation of the electrostatic spectrometer of "Troitsk Nu-mass" experiment

Ivanov B.V.[1], Pantuev V.S.[2], Bukin A.N.[3], Semenov A.A.[3], Belyakov M.I.[3], Belesev A.I.[2], Geraskin E.V.[2], Ionov N.A.[2], Parfenov V.I.[2]

[1] *National Research Center "Kurchatov Institute", 123182, Moscow, Russia*

[2] *Institute for Nuclear Research Russian Academy of Sciences, 117312, Moscow, Russia*

[3] *Bochvar High-Technology Research Institute of Inorganic Materials, 123060, Moscow, Russia*

**Keywords:** tritium, tritium contamination, Troitsk Nu-Mass setup, tritiated water, detritiation, tritium removal

**Abstract**

The paper describes methods and presents results of the "Troitsk Nu-mass" experiment spectrometer cleanup, which inner volume (40 m$^3$) and surfaces (160 m$^2$) was contaminated by 4.4 GBq of tritium. The "Troitsk Nu-mass" experiment of Institute for Nuclear Research of Russian Academy of Sciences, Moscow, is designed to measure the spectrum of electrons from tritium decays in order to search for hypothetical particles - sterile neutrinos. As a result of equipment failure, the spectrometer internal volume was contaminated with tritium. The contamination made measurements impossible and the research program stopped. The methods of vacuum extraction, hydrogen soaks, and water vapour soaks were used for cleanup. As a result of detritiation, the background level of the main detector of the "Troitsk Nu-mass" spectrometer decreased by more than 10 times, which made it possible to resume work. The results are consistent with the data obtained earlier for volumes in normal conditions in the air and can be used for planning work on detritiation of similar installations.

**Introduction**

At present, a large amount of data has been accumulated on the detritiation of various materials and substances used in tritium handling [1][2][3][4][5][6][7], however, the detritiation process is still complicated, resource- and economically expensive procedure in view of the tritium removal processes specifics and the requirements of low levels of residual tritium in materials [8]. This problem is especially important for fusion facilities where huge tritium amounts will be used. For example, JET and TFTR tokamaks used 20 and 25 grams of tritium respectively [9], [10] at the same time, several kilograms of tritium will be used at the ITER reactor [11]. Tritium retention in the fusion reactor materials is one of the most significant reasons for the inefficient use of tritium and safety issues. Therefore, the detritiation of a fusion reactor equipment and materials is an extremely important problem. The investigation of tritium retention in materials and its detritiation is widely investigated in different laboratories [1]–[4], [12], [13]. But the approaches that can be applied for detritiation of large facilities with a relatively low concentration of tritium in materials have not been experimentally verified enough. So, any experimental results on the tritium removal from large scientific facilities are of interest for fusion reactors operation. In this work, detritiation was required to levels of tritium contamination much lower than the normative established ones, which is an even more difficult task. When detritiation of facilities and equipment (as opposed to individual contaminated materials and radioactive waste), additional difficulties are operational limitations of the facility (for example, permissible temperature regimes, pressure ranges, etc.). In case of the "Troitsk Nu-mass" setup, pressure in the spectrometer was limited to 2 mBar and the heating temperature of the outer jacket could not exceed 110° C. These limitations reduce the range of possible methods and lead to additional costs, in particular, an increase in the duration of detritiation.

In this work, we describe the procedure and the results of tritium decontamination in "Troitsk Nu-mass" spectrometer.

**Existing international detritiation experience**

The accumulated experience in removing tritium from materials, including structural materials [1]–[4], is often impossible to directly transfer to detritiation of installations and equipment working with tritium. As an example, the authors of work [1] developed detritiation methods of JET tokamak structural materials. For this purpose, different methods were tested, like, induction vacuum melting, melting in an inert gas and hydrogen atmosphere, static and dynamic isotopic exchange with both molecular hydrogen and humid air, open flame processing. Despite the high efficiency, destructive methods, as well as methods requiring heating to high temperatures used in [1], cannot be applied to the spectrometer detritiation. We decided to use safe procedures with gaseous hydrogen and water vapour.

There is a limited number of experimental works devoted to the detritiation of large-scale installations. Table 1 and Figure 1 summarize the literature data and compare it to the results of this work.

Table 1 Main parameters and conditions of facilities which used detritiation of large volumes.

| Parameter | Type of experiment | | | | |
| --- | --- | --- | --- | --- | --- |
| | INR RAS, Russia | CATS-experiment, Japan [14], [15] | TSTA-facility, USA [16] | TFTR, USA [10] | JET, UK [9] |
| Installation characteristics | Beta Spectrometer | Physical model of a protective shelter for a tritium plant | Protective shelter (room) of tritium plant | Thermonuclear reactor (tokamak) | Thermonuclear reactor (tokamak) |

| Internal volume | 40 m³ | 12 m³ | 2900 m³ | 80 m³ | 200 m³ |
|---|---|---|---|---|---|
| An initial amount of tritium, Bq (Ci) | ~ 5×10⁹ (~ 0,14) | 2,6×10⁹ [1] (0,07) | 5,4×10⁶ (1,5×10⁻⁴) | ~ 6,5×10¹⁴ (~ 17500) | ~ 2,2×10¹⁵ (~ 60000) |
| System pressure range, Bar | $10^{-11} - 2\times10^{-6}$ | ~1 | ~1 | $10^{-6} - 1$ | $10^{-6} - 1$ |
| Methods used | Flushing with various gases during heating | Ventilation with air at different humidity | Air ventilation | Plasma exposure, evacuation with heating, blowing with various gases, etc. | Plasma exposure, evacuation with heating, blowing with various gases, etc. |
| The character of tritium contamination | Adsorption of tritium in the form of a molecule of water and hydrogen from the gas phase | Adsorption of tritium in the form of a molecule of water and hydrogen from the gas phase | Tritium in the form of molecular hydrogen from the gas phase | The incorporation of tritium into materials as a result of plasma irradiation | The incorporation of tritium into materials as a result of plasma irradiation |

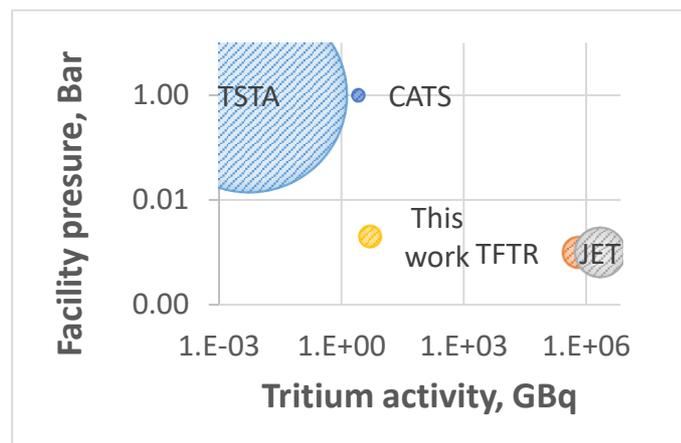

Figure 1. Basic parameters of installations. The size of the circle indicates the internal volume of the system/unit.

Depending on the amount of tritium used during the cleanup procedure, these facilities are divided into 2 groups. The first group includes experiments, like, the TFTR [10] and JET [9] tokamaks, which used a tritium-deuterium mixture and worked at plasma exposure. Some parts were operated in a wide temperature range (from liquid helium temperatures and up to thousands of ºC in plasma). Besides this, they operate at low pressures (up to 100 mBar), with large amounts of tritium (up to 20 g), with additional to plasma loads, including high-energy beams of neutral particles, fusion neutrons, etc. The second group includes experiments of the physical modelling of tritium leakage into a protective room (tertiary confinement) [14], [16]. These studies are characterized by small amounts of tritium, no damaging effects, presence of air at atmospheric pressure in the internal volume, forced ventilation. Detritiation conditions and methods used for these groups differ significantly. "Troitsk Nu-mass" setup occupies an intermediate position between these two groups. It is comparable in size and operational parameters (temperature regimes and pressure ranges) to the first group, but this group uses much larger amounts of tritium and significantly differs in the nature of contamination (plasma irradiation). At the same time, the amount of tritium and the character of contamination are comparable to the second group, but this group differs in the pressure in the system and the absence of heating capability during detritiation. Nevertheless, the earlier experimental results will help to interpret and generalize the results of the current work.

---

[1] This amount of tritium was introduced into the CATS in each experiment

**The "Troitsk Nu-mass" experiment**

The "Troitsk Nu-mass" setup [17] was designed to measure tritium decay spectrum to estimate the electron neutrinos mass and search for hypothetical particles - sterile neutrinos [18]. The setup consists of two main parts: a gaseous windowless tritium source and an electrostatic spectrometer with magnetic collimation, Figure 2. Between the source and the spectrometer, there is a perspectrometer - vacuum-tight volume 0,5 m in diameter and 1 m in length with independent pumping outlet.

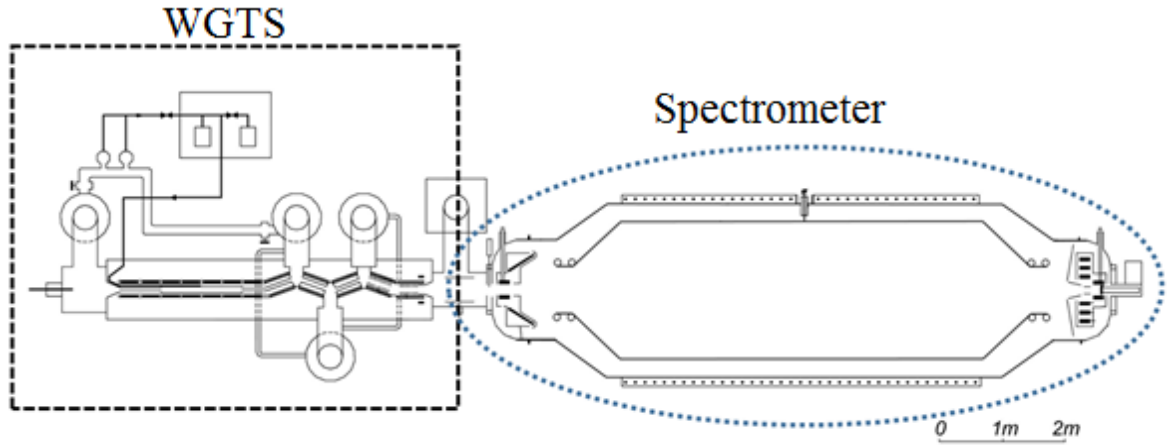

Figure 2. On the left – WGTS, a windowless tritium gaseous source. On the right is an electrostatic spectrometer.

The windowless tritium source, WGTS, (Figure 2) is designed to form an electrons flux generated by tritium decay inside the source tube and transport this flux along magnetic field lines to the spectrometer. In the source, tritium circulates in a closed loop (Figure 3). A booster pump provides the pumping. Tritium from the sorption storage in the form of DT molecules is injected into the centre of decay volume, 'pipe', then gas is pumped out from both sides by pumps P1-P4 into a common collector. Pumps P2, P3, P4 are installed from the spectrometer side and provide differential pumping of the working gas to prevent the tritium penetration into the spectrometer. Additionally, a zigzag-shaped vacuum chamber is installed between the source and the spectrometer. It contains copper cryo-panels with frozen argon trap to absorb tritium. In total, the suppression coefficient of tritium and other gases before entering the spectrometer is not worse than $10^{10}$. At the same time, this does not create obstacles for electrons to enter the spectrometer.

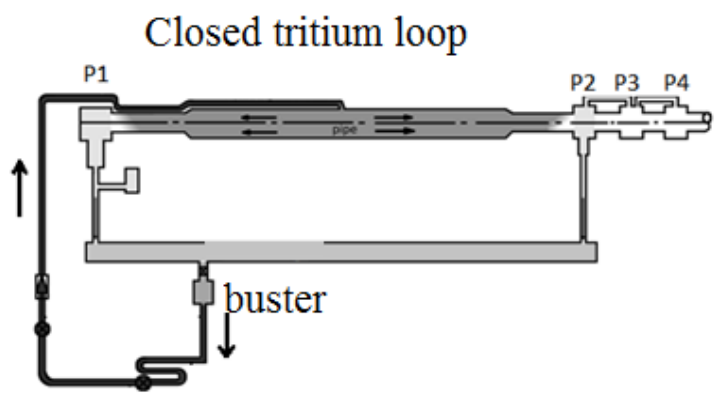

Figure 3. Diagram of a windowless tritium source.

The gas in WGTS sequentially passes through the circuit warm part at room temperature, and then through the central part at a temperature of about 30 K, located inside the superconducting solenoids. At a time about 3 GBq of tritium is injected into the circuit. During measurements, part of the tritium is absorbed on the pipe surface and argon frost. As a result, during long-term operation, there is a decrease in the source intensity by nearly 2 times in 5-6 days. This leads to the need to add tritium to the circuit. Thus, in two weeks of operation of the installation, ~ 4-5 GBq of tritium is absorbed in the circuit cold sections. In addition to tritium, impurity gases from the atmosphere (mainly nitrogen, oxygen) are absorbed over time in the source cooled parts. In the source warm part, where the main part of the working gas is concentrated, the pressure does not exceed 1-2 mBar. A gas mass analyzer controls the gas composition in the circuit warm part between the outlet of the P1 pump and the booster pump. According to records, the partial pressure of water is 5-6 times less than the pressure of nitrogen. After completing work with the source, the vacuum valve between it and the spectrometer is closed. The absence of tritium inside the spectrometer guarantees high-quality measurements of the energy spectrum for electrons generated in WGTS.

After one of the measurement, it was found that the electron background count rate by the main Si-detector located at the end of the spectrometer increased 100 times relative to the previous values and reached 18 000 Hz, which indicates the tritium presence in the spectrometer. This background level was unacceptable for further measurements. Later, it was found that, after the measurement session, the vacuum valve between the source and the spectrometer did not close tightly. This led to tritium penetration into the spectrometer. Between the tritium source and the spectrometer, there is a 0.6 $m^3$ pre-spectrometer, which was also contaminated. It should be noted that most of the tritium was pumped from the cold part of the source and only after that, the pumps (P1-P4 in Figure 3) were stopped. Therefore, only part of the tritium which was adsorbed on the cooled pipes' surfaces and the argon trap could get into the spectrometer. When the source was heated from 30 K to room temperature, defrosting and desorption of impurity gases (nitrogen, argon, oxygen, and water vapour) also occurred. These gases and tritium compounds could penetrate into the spectrometer.

**Spectrometer.**

A spectrometer cross-section is shown in Figure 4. The total spectrometer length is approximately 10 meters and an internal volume is approximately 40 $m^3$. Inside the stainless-steel vessel, there is a central electrode mounted on several dielectric supports. The spectrometer outer diameter is 2.75 m, the cylindrical part length is 6 m, and the inner electrode diameter is 2.4 m. Additional small electrodes and superconducting magnets are installed at the spectrometer ends.

It should be noted that the electrode increases the total surface inside the spectrometer by about 3 times relative to the outer vessel. As a result, the total surface is ~160 m$^2$.

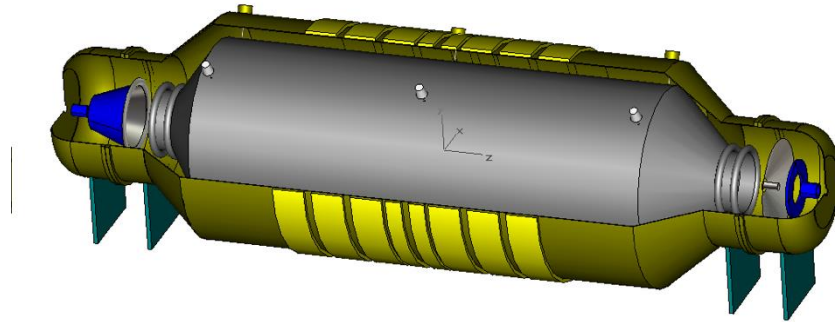

Figure 3. The spectrometer of the "Troitsk Nu-Mass" setup.

**Detritiation condition limitations**

It was decided to perform detritiation of the spectrometer volume without opening it since the filling and pumping cycle of such a volume takes several months. Necessary safety measures were taken to work with tritium and its compounds. Operating parameters and safety considerations limited the maximum pressure (2 mBar) and the temperature of the spectrometer vessel warming up (110°C). The cryostats of superconducting magnets at the input and the end of the spectrometer have indium seals. Heating the spectrometer jacket above 110 ° C can damage the seals and loss cryostats tightness. By limiting the pressure in the spectrometer, it wasn't possible to achieve good heating of the inner electrode surface. The temperature on the electrode surface was not controlled, but calculations show that after 8 hours of warming up, the electrode temperature should reach a value of 60 ° C (see Appendix). In addition, layers of insulation based on aluminized mylar are installed around the cryostats. A thin layer of aluminium is very sensitive to the possible use of ammonia. Viton gaskets, which are used to seal the spectrometer, are also sensitive to ammonia and high temperature.

**Detritiation procedure for contaminated volumes**

Vacuum extraction method, hydrogen and water vapour soaks were used for detritiation.

**Vacuum extraction.** The "Troitsk Nu-mass" vacuum system allows to reach vacuum in the spectrometer at a level of 10$^{-11}$ mBar. This procedure removes tritium from the gas phase and tritium absorbed on the surface. The method has low efficiency at room temperature, since the tritium and its compounds desorption rate is rather low, which is confirmed by the data of detritiation of contaminated samples [2], as well as the tritium removal from the JET and TFTR [9], [10].

**Hydrogen soaks.** In this case, the spectrometer was filled with dry hydrogen to a pressure of $10^{-2}$ mBar and kept for several days without heating the body, and then it was pumped to a pressure of $\sim 1\times 10^{-5}$ mBar. This method was applied for 6 sessions.

**Water vapour soaks.** The spectrometer or perspectrometer volumes were filled with a small portion of hydrogen (protium or deuterium) with the addition of water (except for some cases). 20 ml of distilled water was injected with a syringe through a special gasket, the water instantly evaporated, which was recorded by an increase in pressure. Then, gaseous hydrogen was added to reach pressure in the spectrometer about 2 mBar. The spectrometer vessel was heated to 110 °C in cycles, mainly for 8 hours daytime. Then, after several days of holding, the spectrometer volume was pumped out by a standard vacuum system to a pressure of $\sim 1\times 10^{-5}$ mBar. A liquid nitrogen cold trap carried out the collection of water vapour. The exhaust from the pumping system was directed to a special ventilation system and then to the atmosphere outside the building.

As part of the procedure the exposure duration, the heating duration and the admitted gases composition were varied (in one case, water was replaced by a 10% aqueous solution of ammonia, in the other, by a 3% solution of $H_2O_2$).

The detritiation of the perspectrometer was carried out similar to that described above. Conditions were changed slightly, maximum pressure was increased to about 80 mBar, formed by 20 ml of water, and the rest – by hydrogen (protium or deuterium). The perspectrometer detritiation was performed in parallel and independently of the work on the main spectrometer.

This detritiation procedure with water vapour was applied for 9 sessions for the spectrometer and 6 sessions for the perspectrometer.

**Methodology for assessing the detritiation effectiveness**

The detritiation efficiency was controlled by two methods: by checking the count rate by spectrometer Si-detector and by measuring the collected tritiated water activity.

Water from the spectrometer was collected by a liquid nitrogen cold trap. We did not use catalytic oxidation of molecular tritium (hydrogen) to evaluate its amount in the spectrometer. Therefore, this was an estimate of the tritiated water content only.

The specific activity of the released water was measured by the scintillation method on a Tri-Carb 2810 counter.

Measurement of the spectrometer Si-detector background is an indirect method for assessing detritiation effectiveness. After each session, the spectrometer was pumped to a vacuum of $10^{-5}$ mBar. This vacuum level makes it possible to register electrons that are generated inside the spectrometer. The sources of these electrons are cosmic rays, which knock out delta electrons from the electrode surface, the natural radioactivity of the electrode material, and internal tritium contamination. For measurements, a high negative voltage (from 15 kV to 20 kV) was applied to the electrometer with no magnetic fields. Because even several Gauss magnetic field turns the low-energy electrons back onto the electrode material. The electrons produced on the inner surfaces are accelerated to an energy equal to the voltage on the electrode. Before the tritium contamination of the spectrometer, the Si-detector background count rate was 160–180 Hz. This count level is a good criterion for spectrometer purity. After contamination by tritium, the Si-detector background increased to 18 kHz. Due to the geometric configuration of the additional electrodes and the aperture of the superconducting magnets, the detector registers electrons only from the inner electrode surface. Other contaminated surfaces remain outside the sensitivity range. Measurements with a Si-detector are qualitative, however, they help estimate the degree of decontamination.

**Spectrometer detritiation results and discussion**

The spectrometer detritiation results by hydrogen and water vapour soaks are presented in Table 2.

Table 2. The results of each session of the spectrometer detritiation.

| № | Detector background, kHz | Exposure time, days | Warm-up time, hours | Released activity, MBq | Reagents |
|---|---|---|---|---|---|
| 0 | 18 | 30 | - | - | $H_2$, $10^{-2}$ mBar, Initial background. |
| 1 | 14 | 30 | 80 | 1840 | 20 ml $H_2O$ + $H_2$ |
| 2 | 6.0 | 8 | 28 | 231 | 60 ml $H_2O$ + $H_2$ |
| 3 | 3.95 | 8 | 72 | 137 | 20 ml $H_2O$ + $H_2$ |
| 4 | 2.76 | 4 | 24 | 32 | 20 ml $H_2O$ |
| 5 | 2.44 | 5 | 24 | 22 | 20 ml (97%$H_2O$ + 3%$H_2O_2$) |
| 6 | 2.11 | 12 | 87 | 31 | 20 ml $H_2O$ + $H_2$ |
| 7 | 2.05 | 3 | 12 | 8 | 20 ml (90%$H_2O$ + 10% $NH_3$) |
| 8 | 1.9 | 6 | 48 | 29 | 20 ml $H_2O$ + $D_2$ |
| 9 | 1.86[2] | - | - | - | - |

**The initial tritium contamination** was estimated based on comparison of the released activity in the form of tritiated water and the silicon detector background.

---

[2] residual background

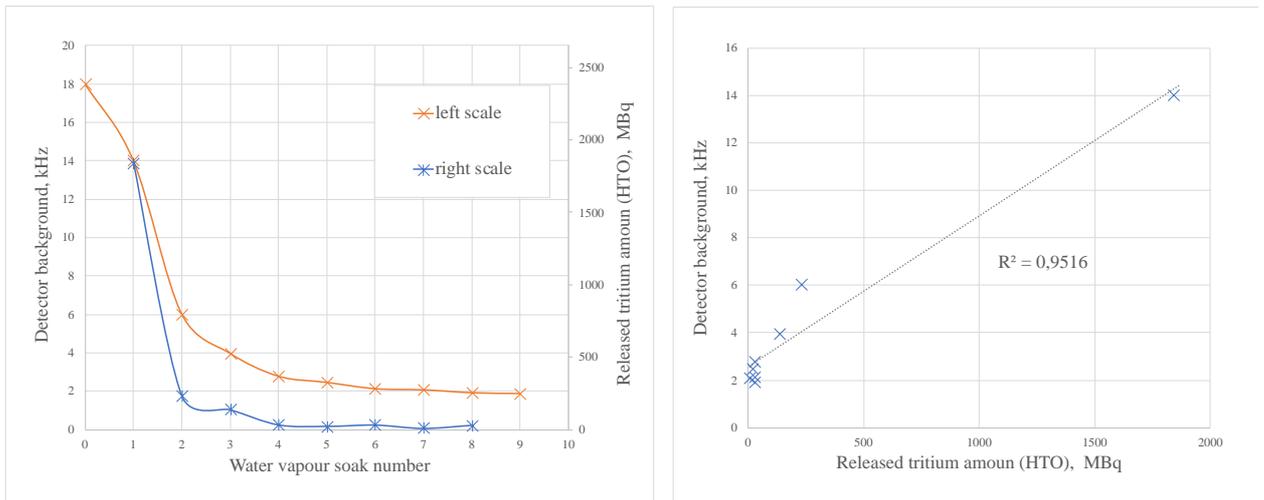

Figure 5. Dependence of the released activity and the detector background on the soaks number (left) and dependence of the detector background on the released activity (right). $R^2$ is a coefficient of determination for linear regression.

Figure 5 shows the correlation between these values. A high residual background of the detector (1.86 kHz) indicates the tritium presence in the spectrometer in such a form (or in such places), the further extraction of which by the selected methods is ineffective. Assuming that the entire decrease in the detector background (from 14 kHz to 1.8 kHz) is a consequence of the removal of 2330 MBq of tritium from the spectrometer, we assume 1 kHz background corresponds to 192 MBq. This leads to estimate initial contamination in the spectrometer at 3.5 GBq (an initial 18 kHz detector background). The total contamination, taking into account the results of cleaning the perspectrometer, is 4.4 GBq ($10^{-5}$ g in terms of pure tritium in a molecular form). It should be noted that the data on the tritium registration efficiency from the gas and absorbed phases by the Si-detector are insufficient; this value error can be significant.

The stabilization of the residual background less than 2 kHz in comparison to the pure spectrometer count rate of 180 Hz indicates that residual tritium content on the inner surfaces at the level of 350 MBq. The specific surface activity of tritium is $\sim$ 200 Bq / $cm^2$ (in case of even spreading over the area of $\sim$160 $m^2$).

**The effectiveness of detritiation methods**

**Vacuum extraction** didn't give significant results. Pumping out the spectrometer and maintaining the vacuum at a level of $1\times10^{-5}$ mBar didn't lead to a decrease in the detector background.

**Hydrogen soaks.** The detritiation process by this method was not monitored at between iterations. Only the initial and final (after 6 cycles) detector backgrounds are available for analysis, 18 kHz and 14 kHz, respectively. These results indicate a rather low efficiency of the method, which also finds confirmation in the literature. The authors of [9] call hydrogen soaks completely ineffective under similar conditions for the JET tokamak, and also refer to similar results for the TFTR [10] and TEX-TOR tokamaks. It should be noted that in tokamaks, almost all tritium is in an absorbed form and is deeply embedded in materials, which is a consequence of the high operating temperature and plasma loads during tokamak operation. In our case, the molecular tritium sorption is possible in a small amount and directly on the material surface only due to low pressure and room temperature. Therefore, seems that it should be easier to remove tritium, but our observation show otherwise.

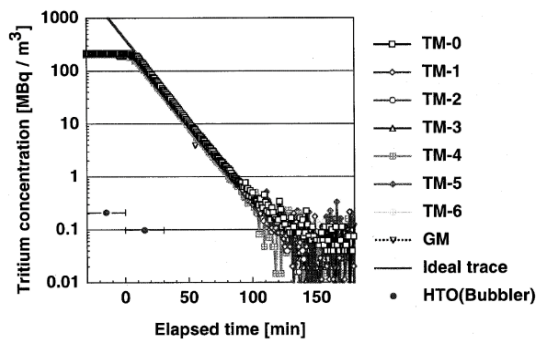

Tritium removal behavior in the dry air atmosphere (humidity:10 ppm) by 50 m3/h of detriation (residence time: 1 week)

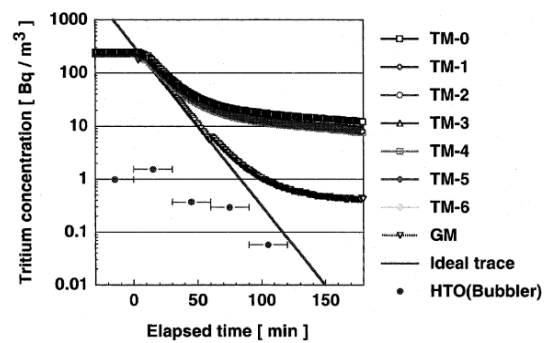

Tritium removal behavior in the dry air atmosphere (humidity:1 ppm) by 50 m3/h of detriation (HTO in the released tritium: 0,4 – 1,0%)

Figure 6. The process of removing tritium from the CATS box with different injected mixture composition. Molecular tritium (left), a mixture of molecular tritium and tritiated water (0.4 - 1.0%) (right) from [14]. Points TM and GM correspond to the readings of various designs ionization chambers installed directly in the box. HTO (Bubbler) points - tritiated water content based on activity data from samples collected with bubblers.

A completely different detritiation behaviour was described in [14], [19]. In these studies, 2.6 GBq of molecular tritium was injected into a special box, then the tritium was removed by a ventilation system. Under such conditions, tritium was completely (to the level of the background of the detectors) removed from the box, on average, with an eight-fold exchange of the atmosphere (Figure 6 left). The authors of [14], [19] especially note that when molecular tritium is injected into a volume of 12 m$^3$ filled with air under normal conditions, there is no significant sorption of molecular tritium by the box walls, as well as the tritiated water formation as a result of oxidation or isotope exchange. That is, the incoming clean air simply displaces tritium from the researched volume.

In our work, there was no displacement, but a dilution of the initial atmosphere inside the spectrometer, the dilution degree was at least 6000 times (6 sessions, the ratio of the initial and final pressure ~1000). As a result, almost all of the tritium in the gas phase was removed. Despite such a significant atmosphere dilution, the detector background decreased by ~20% only. This allows us to conclude that tritium in the spectrometer was distributed between the gas phase and the solid phase (in the absorbed form). After hydrogen soaks, tritium was predominantly in an absorbed form as hydrogen (HT, DT) or water (HTO) molecules.

**Water vapour soaks**. After the first attempt, the spectrometer Si-detector background was significantly reduced. The tritium concentration in the collected water was 92 MBq/ml, which corresponds to a total recovered activity of about 1840 MBq. Subsequently, the background level actually stabilized, and the amount of activity removed decreased significantly, starting with Session 5. Varying of the procedure did not significantly affect the detritiation efficiency. A water amount increase, Session 2, only reduced the specific concentration of tritium per millilitre in recovered water and did not affect decontamination noticeably. Hydrogen peroxide solution 3% was used in Session 5 without adding hydrogen gas. No significant effect was found; rather, a drop in the recovered activity was observed. Ammonia (10%) in water was used in Session 7 to see its potential benefit. For fear of notable corrosion on internal components, the exposure time was limited to three days. Due to the low concentration of ammonia and the short exposure time, the recovered activity dropped by 4 times. Replacing hydrogen (protium) with deuterium, Session 8, also did not improve detritiation. The stable and relatively small amount of tritium in samples 5–8, possibly, indicate that the method limits have been reached. For further detritiation higher temperature (especially on electrode surfaces), pressure and water vapour amount is needed.

The detritiation results of the perspectrometer volume are presented in Table 3. The only way to control the detritiation process of the perspectrometer was to measure the recovered water activity. The detritiation results correspond to described above for the spectrometer. The released tritium amount stabilized after Session 3. The insignificant value in the recovered activity in Session 4, possibly, indicates the importance of the presence of hydrogen during decontamination. A similar effect was observed for Session 4 and 5 from the spectrometer.

Table 3. Results of each session of the perspectrometer detritiation.

| № | Exposure time, days | Warm-up time, hours | Released activity, MBq | Note |
|---|---|---|---|---|
| 1 | 20 | 40 | 426 | 10 ml $H_2O$ |
| 2 | 8 | 32 | 57 | 10 ml $H_2O$ |
| 3 | 9 | 40 | 29 | No hydrogen, 20 ml $H_2O$ |
| 4 | 12 | 64 | 40 | 20 ml $H_2O$ |
| 5 | 18 | 95 | 40 | 20 ml $H_2O$ |
| 6 | 10 | 50 | 30 | 20 ml $H_2O$ |

The first water injection in the spectrometer removed 79% of tritium in the water form (1.8 from 2.3 GBq) and 53% of the total tritium (1.8 from 3.5 GBq), as was estimated above. The Si-detector readings dropped by ~70% after all water vapour soaks, which indicates a high efficiency of the method. This high efficiency should be related to the high rate of transfer of adsorbed tritium to the gaseous phase since tritium was predominantly in the absorbed form before this stage. This process depends mainly on the tritium form (molecular, oxidized or other). The main difficulty to describe the detritiation mechanism is the lack of data on the composition of the gas that entered the spectrometer because of equipment failure. We also don't know the ratio of tritium in various forms (molecular or oxidized or other) during detritiation, since the sampling technique allowed us to capture only tritiated water vapour.

In publications [14], [15], [19] a mechanism of large volume detritiation in the presence of tritiated water is proposed. In these works, in addition to experiments with tritium in molecular form, a mixture of molecular (97-99%) and oxidized (water, up to 3%) tritium with an activity of 2.6 GBq was injected into the box. In the presence of tritiated water, the character of the box detritiation changed significantly. This was revealed in an increase in the time required to evacuate tritium and residual contamination on inner surfaces, the influence of the purged air humidity on the detritiation rate. The authors of articles [14], [15], [19] associate these effects with the process of tritiated water adsorption and desorption on the box inner surfaces. The authors showed that the results of modelling those processes using the Freundlich equation [20] are in good agreement with the experimental results. Figure 6 (right) shows the box detritiation process in the presence of tritiated water in injected tritium. This case resembles in main details the detritiation process we observed.

**Detritiation mechanism.** As noted above, a complete description of the detritiation mechanism needs additional data on the initial contamination and the composition of gases pumped out from the spectrometer and perspectrometer. However, based on a comparison of the data obtained in this work with the results of [14], [15], [19] it is possible to assume the presence of tritiated water at the initial contamination. In this case, tritiated water was easily absorbed on the spectrometer inner surfaces. The high efficiency of water vapour soaks can be explained by the dilution of tritiated water on the surface and its subsequent removal by pumping. The inefficiency of hydrogen soaks is associated with low hydrogen adsorption and low rate of isotope exchange between hydrogen and water.

However, as follows from the description of the tritium source operation, tritiated water should not be there. Therefore, the assumption of its presence at the initial contamination composition may not seem justified and another mechanism should be proposed. In case of only molecular tritium (HT, DT) contamination, a large internal and well-prepared spectrometer surface (almost always under a deep vacuum) can adsorb a significant amount of molecular tritium, even at its low specific content. For example, in [21], a gold-coated sample was exposed in tritium at a pressure of $10^{-5}$ mBar, and the determined surface tritium activity was $5.4 \times 10^4$ Bq/cm$^2$, which significantly higher than our estimation for the spectrometer (~200 Bq/cm$^2$). In this case, the presence of tritium in HTO form in collected water can be explained because of isotope exchange reaction between the adsorbed hydrogen and water. Such a reaction is possible in the presence of materials exhibiting catalytic activity to it, for example, platinum, palladium, nickel, and some others. However, it is difficult to explain the low efficiency of hydrogen soaks, since the adsorbed tritium should be significantly diluted during the exposure and subsequent pumping.

To substantiate the mechanism of detritiation, additional measurements are required, in particular, the analysis of impurities in a windowless tritium source.

**The outcome:**
- "Troitsk Nu-Mass" setup spectrometer and perspectrometer detritiation was carried out. The initial tritium amount is estimated at 4.4 GBq. The residual tritium content in the spectrometer is 0.35 GBq.
- Detritiation was carried out taking into account serious limitations for pressure and temperature in the setup. The minimum amount of reagents was used, which led to a small amount of tritium-containing waste.
- Effectiveness of various detritiation methods was evaluated. The most effective method is the water vapour soaks. An addition of gaseous hydrogen seemed to be effective. It is shown that the addition of ammonia, peroxide and replacement of protium by deuterium does not lead to an increase in the detritiation efficiency. Usage of hydrogen soaks is inefficient.
- The most probable mechanism of spectrometer detritiation is are described.

**Conclusion**

To ensure the "Troitsk Nu-Mass" setup operability in the presence of residual tritium in the spectrometer, it was decided to increase the magnetic field in the spectrometer centre using an additional solenoid winding over the vessel. The additional magnetic field functions as follows. Almost all of the electrons that are formed on the electrode inner surface have extremely low energy, no more than a few electron volts. Even a small magnetic field is enough to turn them around and direct them back into the electrode body. A technical run was carried out with turning on of all elements, including superconducting magnets. A positive result was obtained: after the detritiation and the inclusion of an additional magnetic field, the Si-detector background with an open gas tritium source (without filling with tritium) turned out to be 3-4 times higher than before, but at an acceptable level for measurements.

**Acknowledgements**

The work was supported by the NRC "Kurchatov Institute (16.07.2019 № 1570) and by the Ministry of Science and Higher Education of the Russian Federation under the contract 075-15-2020-778. The authors are grateful to S.V. Suslin for help in this work.

**Appendix. Internal electrode warm-up estimation**

During detritiation, the outer vessel was heated. The transfer of heat inside was mainly through the gas at low pressure. The thermal radiation was at a much lesser extent. The transfer of heat between two surfaces through a gas can be express as follows:

$$Q = -k * \frac{S}{l} * \Delta T,$$

where $k$ is the value of thermal conductivity, $S$ is the area of the heated surface, $l$ is the distance to another surface, $\Delta T$ is the temperature difference. The thermal conductivity of a gas, in turn, at certain pressures is proportional to the product: $k \sim \rho*\lambda*1/\sqrt{m}$, where $\rho$ is the density of the gas, $\lambda$ the length of the free path, $m$ is the molar mass of the gas. Thermal conductivity of air versus pressure is shown in Figure 7.

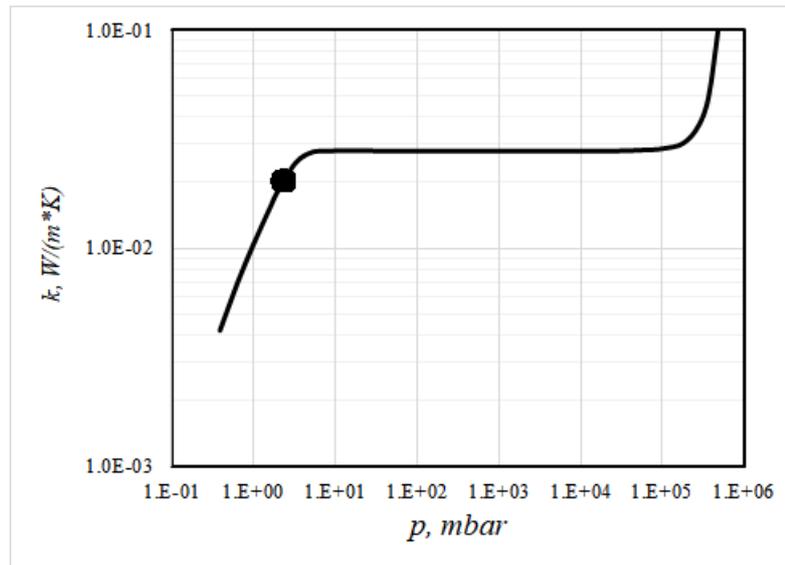

Figure 7. Dependence of air thermal conductivity on pressure. The black square symbol marks the operating mode with about 2 mBar pressure.

The constancy of thermal conductivity in a wide range of pressures is explained by the presence of the product «$\rho*\lambda$» in the previous formula, which compensates for the change in the pressure. Under high vacuum conditions, the coefficient decreases approximately with the density of the gas or with pressure. It should also be noted, according to the formula, due to the dependence of *m*, the thermal conductivity is higher for light gases, in particular, hydrogen has a thermal conductivity 2.6 times higher than that of air. The choice of a filling pressure of no more than 2 mBar is primarily due to the possibility of faster pumping to a pressure of better than $10^{-5}$ mBar for subsequent control measurements of the background with a silicon detector. On the other hand, the filling was determined by the finite amount of hydrogen and water that could be filled in at a time. Taking into account the thermal conductivity mainly through hydrogen with a coefficient of $k = 5.28 * 10^{-2}$, the geometry of the spectrometer with a gap between the body and the electrode $l = 0.2$ m, the heating area of the outer jacket in $S = 47$ m² and the difference in initial temperatures between the vessel and the electrode of the order of $\Delta T = T_o - T = 90$ degrees, where $T_o$ is the temperature of the heated ~~case~~ surface, T is the temperature of the electrode, an estimate of the initial value of heat transfer of the order of 1 kW is obtained. The differential equation for increasing the temperature T of the inner electrode with time *t* is written as:

$$\frac{dT}{dt} = \frac{kS}{lqm}(T0 - T),$$

where *m* is the mass of the electrode, *q* is the specific heat of the material. Solving this equation, we come to the form:

$$T(t) = T0 - DT \exp\left(-\frac{t}{B}\right), \text{ where the time constant is } B = {lqm}/{kS}.$$

With a total electrode mass of about $m = 1000$ kg and a specific heat capacity stainless steel about $q = 500$ J/(kg×deg), the value of B is ~ $4\times10^4$ sec or ~ 12 hours. Heating for 8 hours should raise the electrode temperature to $60^\circ$C.